\documentclass[conference]{IEEEtran}

\IEEEoverridecommandlockouts
\usepackage{cite}
\usepackage{amsmath,amssymb,amsfonts}
\usepackage{algorithmic}
\usepackage{graphicx}
\usepackage{textcomp}
\usepackage{xcolor}
\usepackage{soul}

\def\BibTeX{{\rm B\kern-.05em{\sc i\kern-.025em b}\kern-.08em
    T\kern-.1667em\lower.7ex\hbox{E}\kern-.125emX}}
\begin{document}

\makeatletter
\newcommand{\linebreakand}{%
  \end{@IEEEauthorhalign}
  \hfill\mbox{}\par
  \mbox{}\hfill\begin{@IEEEauthorhalign}
}
\makeatother

\title{Health Guardian Platform: A technology stack to accelerate discovery in Digital Health research\\
}

\author{\IEEEauthorblockN{Bo Wen}
\IEEEauthorblockA{\textit{Digital Health} \\
\textit{IBM T.J. Watson Research Center}\\
Yorktown Heights, USA \\
bwen@us.ibm.com}
\and
\IEEEauthorblockN{Vince S. Siu}
\IEEEauthorblockA{\textit{Digital Health} \\
\textit{IBM T.J. Watson Research Center}\\
Yorktown Heights, USA \\
vssiu@us.ibm.com}
\and
\IEEEauthorblockN{Italo Buleje}
\IEEEauthorblockA{\textit{Digital Health} \\
\textit{IBM T.J. Watson Research Center}\\
Yorktown Heights, USA \\
ibuleje@us.ibm.com}
\linebreakand
\IEEEauthorblockN{Kuan Yu Hsieh}
\IEEEauthorblockA{\textit{Digital Health} \\
\textit{IBM T.J. Watson Research Center}\\
Yorktown Heights, USA \\
kuan.yu@ibm.com
\and
\IEEEauthorblockN{Takashi Itoh}
\IEEEauthorblockA{\textit{Digital Health} \\
\textit{IBM Research}\\
Tokyo, Japan \\
jl03313@jp.ibm.com}
\and
\IEEEauthorblockN{Lukas Zimmerli}
\IEEEauthorblockA{\textit{Digital Health} \\
\textit{IBM Research}\\
Zurich, CH \\
lukas.zimmerli@gmail.com}
\linebreakand
\IEEEauthorblockN{Nigel Hinds}
\IEEEauthorblockA{\textit{Emerging Technology Engineering} \\
\textit{IBM T.J. Watson Research Center}\\
Yorktown Heights, USA \\
nhinds@us.ibm.com}
\and
\IEEEauthorblockN{Elif Eyig\"oz}
\IEEEauthorblockA{\textit{Center for Computational Health} \\
\textit{IBM T.J. Watson Research Center}\\
Yorktown Heights, USA \\
ekeyigoz@us.ibm.com}
\and
\IEEEauthorblockN{Bing Dang}
\IEEEauthorblockA{\textit{Digital Health} \\
\textit{IBM T.J. Watson Research Center}\\
Yorktown Heights, USA \\
dangbing@us.ibm.com}
\linebreakand
\IEEEauthorblockN{Stefan von Cavallar}
\IEEEauthorblockA{\textit{Orygen Digital}\\
Parkville, Victoria, Australia \\
svcavallar@hotmail.com}
\and
\IEEEauthorblockN{Jeffrey L. Rogers}
\IEEEauthorblockA{\textit{Digital Health} \\
\textit{IBM T.J. Watson Research Center}\\
Yorktown Heights, USA \\
jeffrogers@us.ibm.com}
}
}

\IEEEoverridecommandlockouts
\IEEEpubid{\makebox[\columnwidth]{\begin{minipage}{\columnwidth} \copyright2022 IEEE. Personal use of this material is permitted. Permission from IEEE must be obtained for all other uses, in any current or future media, including reprinting/republishing this material for advertising or promotional purposes, creating new collective works, for resale or redistribution to servers or lists, or reuse of any copyrighted component of this work in other works.
\end{minipage}} \hspace{\columnsep}\makebox[\columnwidth]{ }}

\maketitle
\IEEEpeerreviewmaketitle

\IEEEpubidadjcol

\begin{abstract}
This paper highlights the design philosophy and architecture of the Health Guardian, a platform developed by the IBM Digital Health team to accelerate discoveries of new digital biomarkers and development of digital health technologies. The Health Guardian allows for rapid translation of artificial intelligence (AI) research into cloud-based microservices that can be tested with data from clinical cohorts to understand disease and enable early prevention. The platform can be connected to mobile applications, wearables, or Internet of things (IoT) devices to collect health-related data into a secure database. When the analytics are created, the researchers can containerize and deploy their code on the cloud using pre-defined templates, and validate the models using the data collected from one or more sensing devices. The Health Guardian platform currently supports time-series, text, audio, and video inputs with 70+ analytic capabilities and is used for non-commercial scientific research. We provide an example of the Alzheimer's disease (AD) assessment microservice which uses AI methods to extract linguistic features from audio recordings to evaluate an individual's mini-mental state, the likelihood of having AD, and to predict the onset of AD before turning the age of 85. Today, IBM research teams across the globe use the Health Guardian internally as a test bed for early-stage research ideas, and externally with collaborators to support and enhance AI model development and clinical study efforts. 
\end{abstract}


\section{Introduction}
Digital Health is an interdisciplinary field gaining momentum in recent years. The ubiquitous presence of mobile and Internet of things (IoT) technologies, the availability of wearable sensors, and the cost-effectiveness of cloud computing services have ushered in a new era of healthcare services focused on predictive, preventative, and personalized care. By bringing the power of new information technology such as edge computing, cloud computing, and artificial intelligence (AI) into healthcare, many traditional practices can be improved, and new methodologies developed.

\begin{figure}[htbp]
\centering
\includegraphics[width=0.9\linewidth]{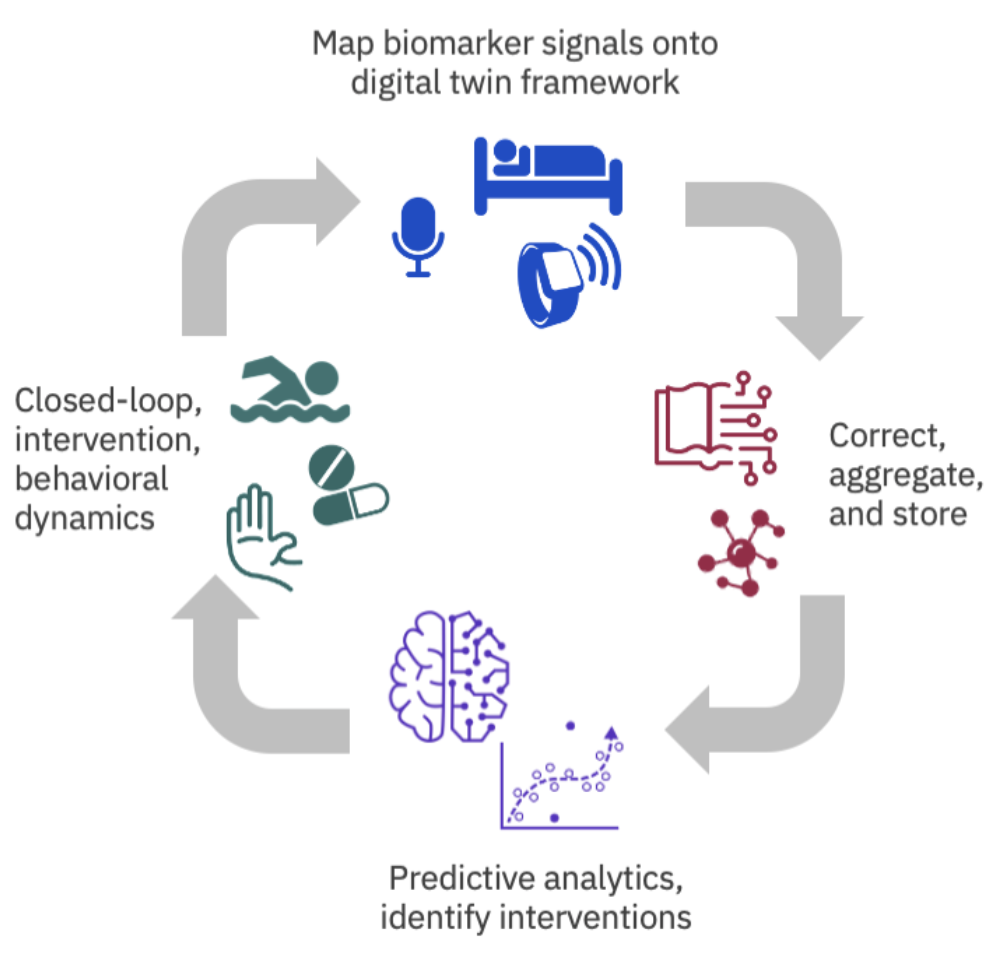}
\caption{An information flow diagram for real-time, remote monitoring of physiological and physical parameters and health insight generation.}
\label{Figure1}
\end{figure}
    
\begin{figure*}[t]
    \centering
        \includegraphics[width=0.73\linewidth]{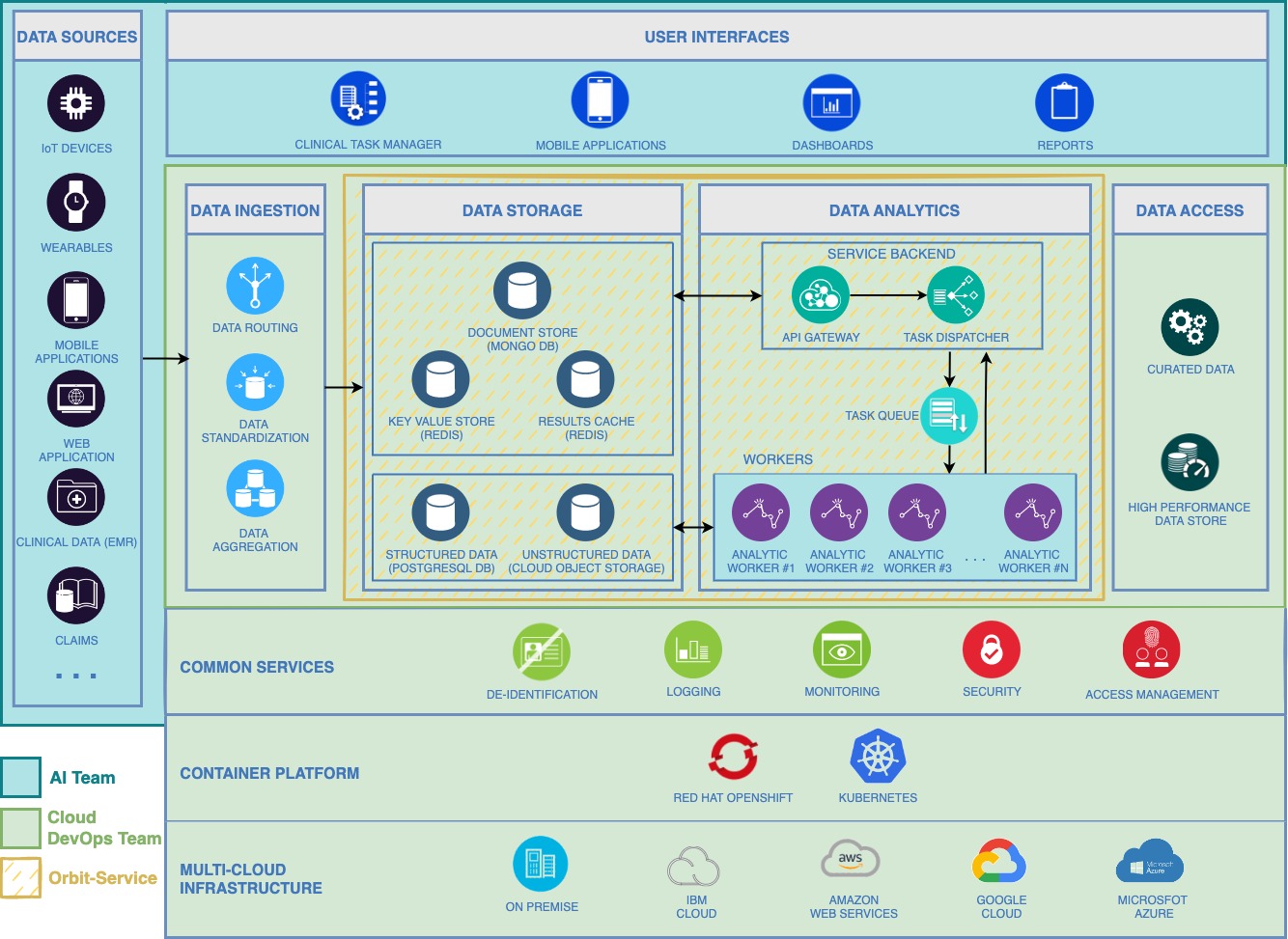}
    \caption{Architecture diagram for the Health Guardian Platform. This schematic is color-coded to denote responsibilities of the AI researchers (cyan), and the Cloud DevOps team (green). The data pipeline consists of four key stages: data ingestion, data storage, data analytics, and data access. The service mesh, Orbit, is denoted in yellow dashes. All components rest on top of a common services layer, and are built on an Openshift and multi-cloud infrastructure.}
    \label{Figure2}
\end{figure*}

For example, the most common form of traditional preventative care are the regular health check-ups, which aims to find potential health issues before they become a problem. However, these exams provide only an episodic view into an individual’s health trajectory, and the infrequency of these check-ups may cause missed windows of optimum intervention. With the combination of wearable and ambient sensing technologies, it is now possible to continuously monitor digital biomarkers \cite{Vasudevan2022} in real-time during activities of daily life. The bio-signals from various smart devices can be transmitted to remote data centers using cloud computing resources to build a holistic model of an individual’s health, or a “digital twin” \cite{Schwartz2020}. The data can be aggregated, processed, and analyzed with AI and machine learning models to identify key health insights and interventions, which would then be relayed back to the patient and their healthcare team. The feedback mechanism and timing can be augmented with modern control theory approaches and behavioral dynamics (Figure \ref{Figure1}).

There are many examples of AI research spanning vast domains of healthcare such as using automated speech and language analysis to identify individuals with Alzheimer's disease \cite{Fraser2015} or determine severity of cognitive impairments \cite{Eyigoz2018,Roark_2007}, automated drawing analysis to assess elderly cognition \cite{Yamada2022}, and gait and motion monitoring for Parkinson's disease assessment \cite{Belic2019}. While many of these algorithms show significant promise, it takes a great amount of coordination to test, revise, and validate the models and successfully translate them into clinical practice. Existing solutions for remote patient monitoring systems are often fragmented and focus on standalone issues of data collection or analysis. Few offer a comprehensive end-to-end solution that provide quality control of every step in the data life cycle: from data collection to analysis to clinical feedback \cite{Sambasivan2021}. 

The Health Guardian described in this paper is a platform that allows for rapid translation of AI research into microservices that can be tested in clinical cohorts. The platform offers an easy-to-use, end-to-end solution for data collection, storage, and analysis, which helps to bridge the gap in translating basic AI research into commercial and clinical use. In this paper, we will describe the Alzheimer's disease (AD) assessment microservice in greater detail. This microservice takes advantage of the data pipeline construction capability of the Health Guardian platform by processing an audio description of a picture displayed on a mobile phone in three main steps. The first two steps uses shared components that involve text pre-processing and natural language processing (NLP) feature engineering algorithms. The final stage can be a regression step to evaluate an individual's mini mental state, or a classification step to compute the probability of them having AD today, or a service to predict the onset of AD before turning 85. \cite{Eyigoz2018,Eyigoz2020}.

The Health Guardian platform is implemented in a hybrid cloud infrastructure to ensure cost-savings in the early phases of a project, and offers scalable solutions to support larger clinical trials and data collection. Standardized and reusable templates are available for new users to easily integrate their analytics into the cloud-based environment. The platform is designed using the “separation of concerns” principle where cloud components are separate from the analytics, so the researchers do not need to be well versed in cloud computing in order to benefit from latest technological advancements. Having a shared and reusable infrastructure enables agile development and streamlines the resources needed to set up new data pipelines for each new project.

\section{Health Guardian Platform}

The Health Guardian platform is a collection of components that can be combined to build custom end-to-end data pipelines that move data through four key stages: data ingestion, data storage, data analytics, and data access. These stages rest on top of a common services layer, and all components are built on an Openshift and multi-cloud infrastructure (Figure \ref{Figure2}).

Data can be ingested from various data sources such as mobile and IoT devices, wearables or electronic health records (EHR). Various data connectors are used to route the data to appropriate datastores. For example, streaming data from IoT devices can utilize Message Queuing Telemetry Transport (MQTT) and Kafka brokers, questionnaire responses from the mobile phone can use HTTP API gateways, and large number of existing EHR files can be processed through open database connectivity/Java database connectivity (ODBC/JDBC) or file exchange systems.

All raw data such as text, audio, images, or video will be separated into structured and unstructured data and stored in a data warehouse or data lake, respectively. Structured data is converted to an IBM proprietary Common Data Model for internal sharing use, and stored in a SQL database. Unstructured data which includes audio, video, and sensor data are stored in its native format in a cloud object storage. Hyperlink to the unstructured data storage location is stored in the database to maintain the association between the datasets. 

The centerpiece of the Health Guardian is an IBM proprietary service mesh system called \emph{Orbit}. This service utilizes the open source Python Celery in a sidecar proxy manner, combined with an IBM proprietary Golang based server for worker registration, worker health monitoring and service discovery. The \emph{Orbit-Service} decouples each analytic service into an API gateway and an analytic worker. The gateway is a light-weight flask application that serves HTTP RESTful API to communicate with the data source or data collecting devices. The analytic worker is created for each microservice and performs the computation analytics on the data collected. 

When a dataset is ingested, the API gateway is invoked by the caller through a batch job script or an event-driven front end application. The API gateway will submit a request to the Task Dispatcher along with the dataset and the task is added to the Task Queue. The next available worker will claim the task and return a task id to the caller. When the worker finishes the computation, the results are saved in the result store with the task id as the key. The caller can then retrieve the result.

On top of the Orbit-based orchestration layer, there is a user interaction application called the \emph{Clinical Task Manager}. Researchers can use this tool to define a clinical study which includes one or more sub-tasks (e.g. questionnaires, audio/video recordings, etc.) for data collection, and assign a cohort to perform those tasks. The tasks will be pushed to the cohort's edge devices (e.g. a smartphone) based on either a schedule set by the researchers, or an event-driven condition. When a subject completes the tasks, the data will be collected by Health Guardian automatically and the \emph{Orbit-Service} will orchestrate the analytics work flow to generate results.

All the components described in the Health Guardian are containerized and can be deployed to any public or private cloud. This allows for a flexible deployment strategy to comply with various privacy and security rules and regulations. There is also a common services layer that includes logging, access controls, and other security measures to ensure data quality and security are maintained throughout the entire data pipeline. 

\begin{figure*}[t]
    \centering
        \includegraphics[width=0.75\linewidth]{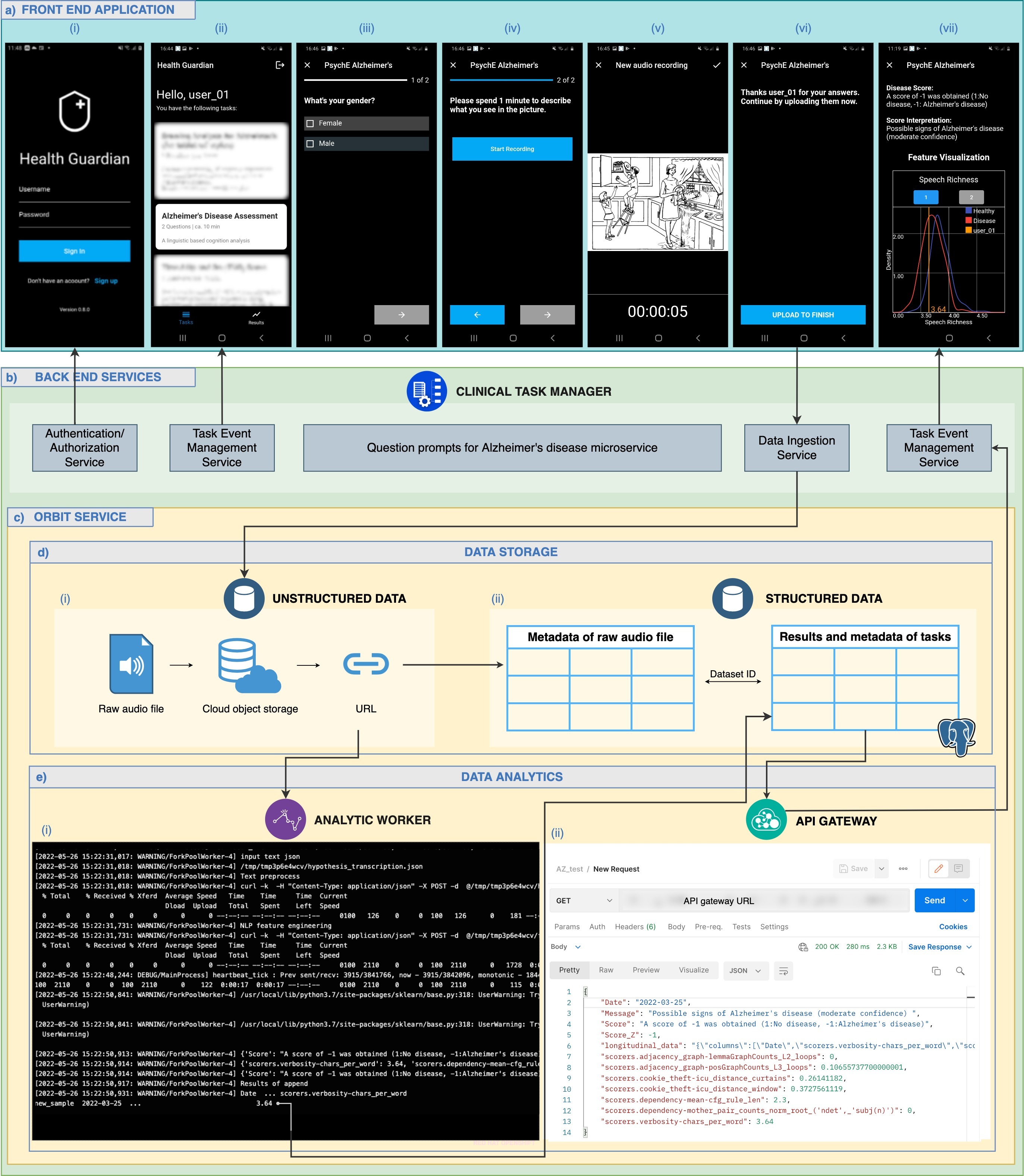}
    \caption{a) Health Guardian front-end mobile application wire frames for the Alzheimer's Disease Assessment task. The Health Guardian back-end components consists of the (b) Clinical Task Manager, (c) Orbit-Service, (d) Data Storage, and (e) Data Analytics. The black arrows in the diagram shows the data piepline between the front-end mobile application to various services in the back-end. The cookie-theft picture description is shown in panel (a.v) captures the audio response of the user, which is then uploaded to the back end (a.vi). The raw audio file is ingested into the cloud object storage (COS) (d.i) and its metadata stored in a raw data table in a postgreSQL database (d.ii). The URL for the raw data file in the COS is sent to the analytic worker to perform the analytics of the  microservice (e.i). When the worker completes the computation, the results are stored in a results table in the postgreSQL database (d.ii). The results can also be retrieved via the API gateway (e.ii) and displayed on the results screen of the mobile application (a.vii).}
    \label{Figure3}
\end{figure*}

\section{Alzheimer's Disease Assessment Microservice}

The Alzheimer's disease (AD) assessment microservice deployed on the Health Guardian platform is an example where AI methods are used to evaluate an individual's mini mental state, the likelihood of AD, and to predict the onset of AD before turning 85 through automated linguistic analysis. The deployment of the microservice onto the Health Guardian platform allows clinical data to be collected on a weekly or daily basis for testing, revising, and validating the computational models. Once the AI model has been validated, the back-end workers and data pipeline of the Health Guardian platform can be used to batch process large clinical datasets in a high-throughput manner. In this section, we will walk through the integration of the AD assessment microservice into the Health Guardian platform.

\subsection{Background}

Alzheimer's disease is a neuro-degenerative disorder that is characterized by a decline in memory, thinking, and independence in activities of daily living. It is the main cause of dementia, and its prevalence is expected to increase from 50 million people in 2010 to 113 million by 2050 worldwide \cite{Knopman2021}. Alzheimer's is a progressive disease, where dementia symptoms worsen gradually over a number of years. The earliest manifestations can be short-term memory difficulty, while during the later stages, individuals can experience impairment in expressive speech, visuospatial processing, and an inability to respond to their environment. 

Besides slowness in thinking and difficulties sustaining attention, language comprehension and production deficits have been well documented in a large variety of cognitive and neurological disorders such as Parkinson's disease, Alzheimer's disease, and aging-related cognitive decline \cite{Lyons1994, Bocanegra2015}. As such, speech and language competence are sensitive indicators for cognitive impairment, and linguistic performance is considered to be a biomarker for neurological disorders. 

In clinical practice, it is common to use a neuro-psychological test such as the cookie-theft picture description task from the Boston Aphasia Diagnostic Examination to evaluate linguistic performance \cite{Goodglass2000, Cummings2019}. The cookie-theft image is one of the most commonly used picture description task and requires an individual to describe what they see in the picture either orally or in written form. 

Many AI researchers have developed computational techniques to extract the linguistic features from speech and text responses. Examples of linguistic variables include verbosity, lexical richness, and repetitiveness. These can be determined by counting the number of words, number of unique words, and frequency of repeated words used. Other linguistic analyses include semantic analysis, which measures how often the text contains information content units relevant to the cookie-theft picture \cite{Giles1996}, and syntactic complexity, a measure of the diversity, variety, and degree of linguistic elaborateness used \cite{Kokkinakis2020}. 

\subsection {Integration into the Health Guardian Platform}

The Health Guardian front-end mobile application is built with a modular architecture. It is capable of showing various types of prompts to interact with the study participants and can capture video, audio, text and other forms of data based on the researcher's experimental design.

To conduct an Alzheimer's assessment experiment, researchers use the \emph{Clinical Task Manager} (CTM) to create an experiment by defining a task with two question prompts. The first prompt asks about the participant's gender, which is useful in the acoustic analysis where vocal features is dependent on the gender. The second prompt asks for a one minute audio recording to describe the cookie theft picture. The researchers can then assign this task to a cohort of select study participants. All participants in a given cohort will receive the task automatically in the Health Guardian mobile application installed on their smartphones (Figure 3a.ii). The Health Guardian mobile application  will unpack the task payload based on the researcher's input in the CTM. It will display the first prompt as a gender question with two choice buttons (Figure 3a.iii). The second prompt will show text-based instructions (Figure 3a.iv) followed by an image of the cookie theft picture when the audio recording starts (Figure 3a.v). It is worth noting that the mobile front-end application is designed with many task visualization template options, such as single/multiple choice questions, sliding scale selection, and text/image/audio/video prompts etc. When researchers create the task and associated question prompts in the CTM, they can select the most appropriate response choice. All of these can be done without requiring the researchers to write a single line of code.

Next, in order to process the audio data, a worker and an API gateway for the AD assessment microservice also need to be created using pre-defined, easy-to-use templates during the experiment design stage:

In the worker template, the researchers need to update all Python package requirements and docker build instructions. Complicated analytic code can be installed as a Python package or imported as a git submodule. Simple analytic code can be embedded directly into the worker template. The entry points for feeding input data to the analytic process need to be defined as Celery jobs and registered in the \emph{Orbit-Service}. Jobs for different analytic processes can be organized into different namespaces in the \emph{Orbit-Service}. Once the worker template is updated with the changes above, it can be containerized and deployed onto any computing infrastructure. In our case, we used an Openshift cluster for production deployment. It is also possible to deploy the workers on Virtual Machine servers, bare-metal servers or even a researcher's personal laptop during the development and testing phases. If the analytic code requires special hardware for computation, like graphic processing units (GPUs) or a quantum computer, the worker can be deployed on such machines as well.

The second template to set up is the API gateway. The API gateway is a light-weight Flask application that exposes callable HTTP endpoints for invoking the jobs registered in the \emph{Orbit-Service}. When researchers create the experiment in the CTM, they will specify the appropriate endpoints for data processing. When the \emph{data ingress service} receives the upload package from the Health Guardian mobile application, it will automatically call the API gateway to submit a job to the appropriate data processing queue. The mobile application can also call the API gateway to retrieve the results when a worker has completed its assigned tasks. All workers and API gateways adopt stringent security requirements including activity logging, access controls, and periodic vulnerability scanning. 

The AD assessment pipeline consists of several stages: the analytics first converts the recorded audio to text using IBM's Speech-to-Text service, and then passes the transcription of the audio file to a text pre-processing service. The text pre-processing service in turn generates representations of text, such as lemmas, syntactic trees, semantic annotations, to be used for further linguistic analyses. These representations are then passed to the natural language processing (NLP) feature engineering service to be filtered down based on the final scoring task's requirement. A final scoring stage will apply classification models to the selected features to calculate a confidence score for the likelihood of AD based on the linguistic features extracted. By simply replacing the final scoring stage with regression models and selecting a different subset of extracted linguistic features, a new service is created to provide a snapshot of an individual's mini mental state. Likewise, a service for predicting the onset of AD before turning 85 can be created\cite{Eyigoz2020}.

When the user completes their description task, the audio recording will be uploaded to the Health Guardian \emph{data ingress service} (Figure 3a.vi). The raw audio file is ingested and stored in the cloud object storage (COS) (Figure 3d.i), while the metadata is stored in a raw data table in a postgreSQL database (Figure 3d.ii). Metadata includes the original task description, data collection details such as the timestamp when the participant started and completed the tasks, answers to the gender question, and url for retrieving the raw audio file. The \emph{data ingress service} will use the data handler information which is setup by the researchers during the experiment design phase to determine which API gateway endpoints to invoke. The process request with the data payload is then sent to the API gateway, the \emph{Orbit-Service} will add the request to the Task Queue, and a \emph{taskID} will be returned. The Health Guardian mobile application can use this \emph{taskID} to retrieve the final results after all the analytics are completed. The task will be assigned to the next available AD assessment worker and the computation performed. An example of the worker logs for a given AD assessment task is shown in Figure 3e.i. The results of the worker are persisted in a results table in the same postgreSQL database as the raw data table. The data from the raw and results data tables are correlated and can be queried using the dataset ID for a specific task performed. The results table is also accessible by the API gateway. The mobile application is set up to ping the API gateway every 10 seconds until a result has been posted (Figure 3e.ii). The mobile application will then parse the results from the API gateway, and recreate a more digestible form to display on the mobile screen. Figure 3a.vii shows an example of the modified results from the API gateway. A confidence score between -1 and 1 is shown, where -1 is the highest likelihood of AD while 1 is the lowest. It also displays a feature graph for speech richness, an example linguistic feature that was extracted. On the graph, the user's performance (orange line) in that feature category is shown in comparison to a distribution of healthy individuals (blue line) and those diagnosed with AD (red line).

\section{Discussion}
The architecture of the Health Guardian platform integrates the agile development methodology \cite{Highsmith2001} into the traditional clinical study workflow. The platform is designed with a reusable infrastructure that establishes a standard communication protocol and data format, so code developed in one project can be leveraged by other projects.

A novel feature of the Health Guardian platform is the decoupling of the analytic service into a worker and API gateway in the \emph{Orbit-Service}. This decoupling offers several key benefits. First, the decoupling optimizes the use of computation resources such that a specialized machine does not waste resources handling common HTTP RESTful requests, nor do web servers or cloud environments have to handle special computing tasks like quantum processing. Second, since the workers and gateways communicate through Python Celery's distributed task queue, researchers do not need to handle time-consuming tasks such as opening ports in remote firewall, managing IP tables for service discovery, or setting up a SSH jumping server through a client's security management team. Third, the light-weight gateway gives flexibility to the data collection since it can be deployed to any public or private cloud. Lastly, this decoupling creates a sand-box environment for researchers to efficiently work from their own laptop (or specialized lab computing environment), while receiving all the benefits of cloud computing: having access to wider range of clinical trial participants without geographical restriction, using elastic scaling services to optimize operation costs, and improving system security and robustness. Since the worker can be deployed anywhere, researchers can run experiments with minimal support from the DevOps team. This reduces the communication overhead, shortens the experiment cycles, and improves overall research efficiency.

In the case with the Alzheimer's disease assessment microservice, multiple workers can be deployed to scale up and expedite batch data processing. If there are multiple projects that require the use of the same worker, and it is desirable to keep the data streams separate, the Health Guardian platform allows for multiple instances of the same workers to be deployed easily by simply changing its name space and where the workers are registered. 

\section{Conclusions}
In recent years, the number of digital health services focused on predictive, preventative, and personalized care have grown tremendously. However, the implementation and adoption of these tools at an enterprise level remains a challenge. The Health Guardian platform and its microservice-based architecture is a tool that can be used to accelerate AI research development, bridging the gap between clinical validation and integration into healthcare workflows. The Health Guardian platform separates the complexity of the cloud infrastructure from other research-related components, allowing researchers to incorporate their models into the cloud infrastructure, and perform iterative cycles of data collection to test and enhance the algorithms using clinical data. Solutions that provide end-to-end data life cycle management, such as the Health Guardian, is key to keep pace and validate the rapid development of new digital health technologies. 

\section*{Acknowledgment}

The authors would like to thank Guillermo Cecchi, Joshua Smith, and Thomas Brunschwiler from IBM for their insights and discussion, and acknowledge support from IBM Research Accelerated Discovery Department.

\bibliography{bibliography}

\begin{thebibliography}{10}
\providecommand{\url}[1]{#1}
\csname url@samestyle\endcsname
\providecommand{\newblock}{\relax}
\providecommand{\bibinfo}[2]{#2}
\providecommand{\BIBentrySTDinterwordspacing}{\spaceskip=0pt\relax}
\providecommand{\BIBentryALTinterwordstretchfactor}{4}
\providecommand{\BIBentryALTinterwordspacing}{\spaceskip=\fontdimen2\font plus
\BIBentryALTinterwordstretchfactor\fontdimen3\font minus
  \fontdimen4\font\relax}
\providecommand{\BIBforeignlanguage}[2]{{%
\expandafter\ifx\csname l@#1\endcsname\relax
\typeout{** WARNING: IEEEtran.bst: No hyphenation pattern has been}%
\typeout{** loaded for the language `#1'. Using the pattern for}%
\typeout{** the default language instead.}%
\else
\language=\csname l@#1\endcsname
\fi
#2}}
\providecommand{\BIBdecl}{\relax}
\BIBdecl

\bibitem{Vasudevan2022}
S.~Vasudevan, A.~Saha, M.~E. Tarver, and B.~Patel, ``Digital biomarkers:
  Convergence of digital health technologies and biomarkers,'' \emph{npj
  Digital Medicine}, vol.~5, no.~1, p.~36, Mar 2022.

\bibitem{Schwartz2020}
S.~M. Schwartz, K.~Wildenhaus, A.~Bucher, and B.~Byrd, ``Digital twins and the
  emerging science of self: Implications for digital health experience design
  and “small” data,'' \emph{Frontiers in Computer Science}, vol.~2, 2020.

\bibitem{Fraser2015}
K.~C. Fraser, J.~A. Meltzer, and F.~Rudzicz, ``Linguistic features identify
  {A}lzheimer’s disease in narrative speech,'' \emph{Journal of
  {A}lzheimer’s Disease}, vol.~49, no.~2, p. 407–422, Oct 2015.

\bibitem{Eyigoz2018}
E.~Eyig\"oz, G.~Cecchi, and R.~Tejwani, ``Predicting cognitive impairments with
  a mobile application,'' in \emph{Proceedings of the 10th International
  Conference on Agents and Artificial Intelligence}.\hskip 1em plus 0.5em minus
  0.4em\relax {SCITEPRESS} - Science and Technology Publications, 2018.

\bibitem{Roark_2007}
B.~Roark, M.~Mitchell, and K.~Hollingshead, ``Syntactic complexity measures for
  detecting mild cognitive impairment,'' in \emph{Proceedings of the Workshop
  on {BioNLP} 2007 Biological, Translational, and Clinical Language Processing
  - {BioNLP}07}.\hskip 1em plus 0.5em minus 0.4em\relax Association for
  Computational Linguistics, 2007.

\bibitem{Yamada2022}
Y.~Yamada, K.~Shinkawa, M.~Kobayashi, D.~G. Varsha D~Badal, E.~E. Lee, R.~Daly,
  E.~W.~T. Camille~Nebeker, C.~Depp, M.~Nemoto, K.~Nemoto, H.-C. Kim, T.~Arai,
  and D.~V. Jeste, ``Automated analysis of drawing process to estimate global
  cognition in older adults: Preliminary international validation on the us and
  japan data sets,'' \emph{JMIR Form Res}, vol.~6, pp. 1--9, May 2022.

\bibitem{Belic2019}
M.~Belić, V.~Bobić, M.~Badža, N.~Šolaja, M.~Durić-Jovičić, and V.~S.
  Kostić, ``Artificial intelligence for assisting diagnostics and assessment
  of parkinson’s disease—a review,'' \emph{Clinical Neurology and
  Neurosurgery}, vol. 184, p. 105442, 2019.

\bibitem{Sambasivan2021}
N.~Sambasivan, S.~Kapania, H.~Highfill, D.~Akrong, P.~K. Paritosh, and
  L.~Aroyo, ``“everyone wants to do the model work, not the data work”:
  Data cascades in high-stakes ai,'' \emph{Proceedings of the 2021 CHI
  Conference on Human Factors in Computing Systems}, 2021.

\bibitem{Eyigoz2020}
E.~Eyigoz, S.~Mathur, M.~Santamaria, G.~Cecchi, and M.~Naylor, ``Linguistic
  markers predict onset of alzheimer's disease,'' \emph{eClinicalMedicine},
  vol.~28, pp. 1--9, 2020.

\bibitem{Knopman2021}
D.~S. Knopman, H.~Amieva, R.~C. Petersen, G.~Chételat, D.~M. Holtzman, B.~T.
  Hyman, R.~A. Nixon, and D.~T. Jones, ``Alzheimer disease,'' \emph{Nature
  Reviews Disease Primers}, vol.~7, pp. 1--21, May 2021.

\bibitem{Lyons1994}
K.~Lyons, S.~Kemper, E.~Labarge, F.~R. Ferraro, D.~Balota, and M.~Storandt,
  ``Oral language and {A}lzheimer's disease: A reduction in syntactic
  complexity,'' \emph{Aging, Neuropsychology, and Cognition}, vol.~1, pp.
  271--281, Dec 1994.

\bibitem{Bocanegra2015}
Y.~Bocanegra, A.~M. Garc{\'\i}a, D.~Pineda, O.~Buritic{\'a}, A.~Villegas,
  F.~Lopera, D.~G{\'o}mez, C.~G{\'o}mez-Arias, J.~F. Cardona, N.~Trujillo
  \emph{et~al.}, ``Syntax, action verbs, action semantics, and object semantics
  in parkinson's disease: Dissociability, progression, and executive
  influences,'' \emph{Cortex}, vol.~69, pp. 237--254, 2015.

\bibitem{Goodglass2000}
H.~Goodglass, E.~Kaplan, and B.~Barresi, \emph{Boston Diagnostic Aphasia
  Examination Record Booklet}.\hskip 1em plus 0.5em minus 0.4em\relax
  Lippincott Williams \& Wilkins, 2000.

\bibitem{Cummings2019}
L.~Cummings, ``Describing the cookie theft picture: Sources of breakdown in
  alzheimer's dementia,'' \emph{Pragmatics and Society}, vol.~10, no.~2, pp.
  151--174, March 2019.

\bibitem{Giles1996}
E.~Giles, K.~Patterson, and J.~R. Hodges, ``Performance on the boston cookie
  theft picture description task in patients with early dementia of the
  alzheimer's type: missing information,'' \emph{Aphasiology}, vol.~10, no.~4,
  pp. 395--408, 1996.

\bibitem{Kokkinakis2020}
\BIBentryALTinterwordspacing
D.~Kokkinakis, K.~L. Fors, C.~Themistocleous, M.~Antonsson, and M.~Eckerström,
  Eds., \emph{Proceedings of the {LREC} 2020 Workshop on Resources and
  Processing of Linguistic, Para-linguistic and Extra-linguistic Data from
  People with Various Forms of Cognitive/Psychiatric/Developmental Impairments,
  RaPID@LREC 2020, Marseille, France, May, 2020}.\hskip 1em plus 0.5em minus
  0.4em\relax European Language Resources Association, 2020. [Online].
  Available:
  \url{https://lrec2020.lrec-conf.org/media/proceedings/Workshops/Books/RaPID3book.pdf}
\BIBentrySTDinterwordspacing

\bibitem{Highsmith2001}
J.~Highsmith and A.~Cockburn, ``Agile software development: the business of
  innovation,'' \emph{Computer}, vol.~34, no.~9, pp. 120--127, 2001.

\end{thebibliography}
\bibliographystyle{IEEEtran}

\end{document}